# Geometric frustration in compositionally modulated ferroelectrics

Narayani Choudhury[1], Laura Walizer[2], Sergey Lisenkov[3], L. Bellaiche[1]

**Geometric frustration is a broad phenomenon that results from an intrinsic incompatibility between some fundamental interactions and the underlying lattice geometry[1-7]. Geometric frustration gives rise to new fundamental phenomena and is known to yield intriguing effects, such as the formation of exotic states like spin ice, spin liquids and spin glasses[1-7]. It has also led to interesting findings of fractional charge quantization and magnetic monopoles[5,6]. Geometric frustration related mechanisms have been proposed to understand the origins of relaxor behavior in some multiferroics, colossal magnetocapacitive coupling and unusual and novel mechanisms of high $T_c$ superconductivity[1-5]. Although geometric frustration has been particularly well studied in magnetic systems in the last 20 years or so, its manifestation in the important class formed by ferroelectric materials (that are compounds exhibiting electric rather than magnetic dipoles) is basically unknown. Here, we show, via the use of a first-principles-based technique, that compositionally graded ferroelectrics possess the characteristic ``fingerprints'' associated with geometric frustration. These systems have a highly degenerate energy surface and exhibit original critical phenomena. They further reveal exotic orderings with novel stripe phases involving complex spatial organization. These stripes display spiral states, topological defects and curvature. Compositionally graded ferroelectrics can thus be considered as the ``missing'' link that brings ferroelectrics into the broad category of materials able to exhibit geometric frustration. Our *ab-initio* calculations allow a deep microscopic insight into this novel geometrically frustrated system.**

Geometrically frustrated systems like spin ice and spin liquids reveal intriguing phenomena and are known to exhibit a degenerate manifold of exotic ground states[1-8]. Spectacular features of geometrically frustrated compounds include formation of complex microstructures[1-8], strong deviation of macroscopic properties from well-known critical behaviors[9-10] and presence of topological defects, spiral states and curvature[11-17]. Here, we report the discovery that compositionally modulated ferroelectrics exhibit all these phenomena that are hallmarks of geometric frustration (GF).

Practically, we investigated compositionally graded $(Ba,Sr)TiO_3$ (BST) compounds, which are systems that are promising candidates for applications as storage capacitors and dielectrics for the next generation of dynamic random access memories[18,19]. We have chosen the saw-like, compositional modulation that is depicted in Fig. 1a and that yields an *overall* Ba and Sr compositions of 50% each. The modulation is periodic along the *z*-axis, that is taken to lie along the *[001]* pseudo-cubic direction (the *x*- and *y*-axes are along the *[100]* and *[010]* directions, respectively). The mimicked modulated system is assumed to be epitaxially grown on a substrate and therefore to adopt the in-plane lattice constant of such substrate. The misfit strain arising from the difference between the in-plane lattice parameter of the compositionally graded material and of the substrate is allowed to vary between -3% and + 3%. (Note that epitaxial strain in oxide perovskite superlattices permits tuning of macroscopic properties and can also result in the formation of novel spatially ordered structures[18-25]). Monte Carlo simulations using effective Hamiltonians[25,26] are performed to predict properties of this strained modulated system for different periodic supercells (see Methods).

Figure 1b reveals that the investigated material has a complex temperature-versus-misfit phase diagram. At high temperature, the phase is the so-called paraelectric, `p' state[22] for all misfit strains, while a large tensile strain yields, at lower temperature, the so-called orthorhombic `aa' state[22] which possesses an in-plane polarization lying along *[110]*. These two phases have also been found in $BaTiO_3$ thin films[23] and $BaTiO_3/SrTiO_3$ superlattices[25]. Similarly, the two other kinds of phase present in the phase diagram of Fig. 1b bear resemblance with some states discovered in some $BaTiO_3/SrTiO_3$ superlattices[25]. They are: (1) `Phase I' that occurs for the largest compressive strains, below the Curie temperature, and that consists of separated `up' and `down' domains, i.e. for which the *z*-component of the electric dipoles is positive and negative, respectively; and (2) `Phase II' which further exhibits an homogeneous in-plane polarization along *[110]*, in addition to the formation of out-of-plane domains, for intermediate strain and below a critical strain-dependent temperature, $T_c$. However, unlike in $BaTiO_3/SrTiO_3$ superlattices[25], the `up' and `down' domains in Phase I and II do *not* ``nicely'' organize into periodic nanostripes alternating along a specific, in–plane direction. The domain states forming Phase I and II are not only rather exotic but are also found to be highly degenerate (see, e.g., Figs. 2): various states with different complex organizations between the `up' and `down' domains, *but with similar energy*, can be found for the same misfit strain (even for the largest studied supercell). Interestingly, anomalous microstructures and ground-state degeneracy are typical signatures of frustration[1-7].

The formation of the exotic structures of Phases I and II is also associated with anomalous dielectric responses. More precisely, the $\chi$ dielectric susceptibility for temperatures, T, above the *p*-to-Phase I and *p*-to-Phase II transitions satisfies the power law $1/\chi \alpha (T-T_c)^\gamma$ over a large range of temperature, with the critical exponent $\gamma$ significantly deviating from its Curie-Weiss value of 1 [see, Figs 1c and 1d]. In particular, $\gamma$ can be as large as 1.8 in Phase I, depending on the resulting microstructure. On the other hand, a similar analysis for the dielectric constant in the vicinity of the *p*-to-*aa* phase transition (i.e., with no out-of-plane domains at low temperature and large tensile strain) yields $\gamma=1$, in agreement with the Curie-Weiss behavior. These dielectric anomalies in Phases I and II are once again consistent with frustration, since GF has been documented to significantly affect critical behaviors[9-10]. We further obtain a correlation between the exponent $\gamma$ and the complexity of the microstructure: the larger the $\gamma$, the more the dipolar configuration differs from the `regular', periodic nanostripe domains previously found in BT films[23] and $BaTiO_3/SrTiO_3$ superlattices[25]. (Note that values of $\gamma$ larger than 1 characterize a phase transition being diffuse in nature; a recent work linked diffuseness with dipolar inhomogeneities[20]). Such correlation is of large benefit to unravel the microscopic initiation and development of frustration in our investigated system. As a matter of fact, Fig. 2(a) (that corresponds to the less frustrated microstructure, since its associated $\gamma$ parameter is the smallest one among all the displayed configurations) reveals that frustration is initiated by flipping of some `up' and `down' dipolar displacements near domain walls. The other panels of Figures 2 demonstrate that frustration further percolates across to other sites, via flipping of dipoles, leading to very exotic shapes for the dipole pattern in the most frustrated systems – such as the ``torus'' or ``cross'' configurations displayed in Figs 2(e) and 2(f). Interestingly, the torus configuration (figs. 2e, 2f) enables us to establish a parallel between our investigated modulated system and geometrically frustrated spin liquids, which are known to have degeneracies on a *torus*[1,8]. Such a parallel in behavior, once again, strongly supports the idea that our studied ferroelectric material is also geometrically frustrated. This idea is further confirmed when realizing that difficulties in precisely predicting physical behavior are a characteristic signature of geometric frustration[1], as consistent with the fact that the numerous, discovered possible ground states (see, e.g., Figs. 2) span a wide range of critical exponents (see, e.g., Fig. 1b and its captions).

Interestingly and as revealed by Figs. 1e and 1f, a common feature emerges between all the various, different-in-morphology states forming Phase I and II: the average magnitude of the *z-component* of the (Ti-centered) local electric dipoles *smoothly* varies as function of the index of the *(001)* $TiO_2$ layer. This magnitude is maximum when the layer is close to a pure BaO plane and minimum when the $TiO_2$ layer is close to a pure SrO plane, as consistent with the fact that decreasing the Ba composition in BST solid solutions results in a smaller polarization at low temperature and to a smaller equilibrium lattice constant. Note that neither Phase I nor Phase II exhibits an out-of-plane component of the overall polarization, as a result of the cancellation of the *z*-component of the dipoles between the `up' and `down' domains.

It was also important to determine if any non-trivial microscopic ordering between dipoles occurred in the frustrated states. Figure 3 (a) displays a three-dimensional view of the dipolar arrangement of a ground state belonging to Phase II, while Fig. 3(b) shows this dipole pattern in a given *(010)* plane, repeating twice along both the *x* and *z*- directions. Let a ``single *z*-chain'' denote a chain oriented along the *z*-direction and passing through all the *(001)* B-layers of the graded system. Figure 3(c) reveals the average angle between dipoles belonging to the same *z*-chain as a function of the distance between the centers of these dipoles. One can see from Figs. 3a and b that most single *z*-chains have either a positive or negative *z*-component of the dipoles (that is, the *z*-component of the dipoles ``only'' alters in *size* from one *(001)* B-layer to another within a *z*-chain). On the other hand, the *in-plane* component of the dipoles is found to significantly vary *in direction* when going from one *(001)* B-layer to another *(001)* B-layer along a given *z*-chain. Such variation is reminiscent of the spiral or helical states that have been reported for other GF systems[16]. This practically leads to an average angle between adjacent dipoles along any up (or down) *z*-chain being close to 10 degree (Fig. 3c). Regarding the overall organization between *z*-chains, Figs. 3a and b reveal that, within a given (001) B-layer, the dipoles can change the sign for their *z*-dipoles when scanning the different *z*-chains along the *x*- or *y*-direction. Several `down' *z*-chains (i.e. with negative *z*-component for their dipoles) exist next to each other and then alternate with successive `up' *z*-chains along the *x*- or *y*- direction. To better appreciate such organization, Fig. 3(d) displays the correlated angle between a dipole belonging to a *z*-chain and another dipole located along the same [100] pseudo-cubic direction as a function of the distance between the centers of these two dipoles, for two different (001) B-layers: one close to a pure BaO plane and one close to a pure SrO plane. Similarly, Figure 3(e) shows the angle formed by two dipoles at the domain walls as a function of the (001) B-layer index. Interestingly, Figs.3(b, d, e) indicate that any (domain wall) connection between ``down'' and ``up'' *z*-chains along the *x*- or *y*-axis involves a complex, cooperative dipolar arrangement. This is evidenced by the fact that the average angle between any two dipoles being adjacent along the [100] pseudo-cubic direction and belonging to up and down *z*-chains, respectively, strongly depends on the position of the (001) B-layer (see Fig. 3e). For instance, this angle is around $150^o$ for (001) B-layers close to pure BaO planes *versus* $60^o$ for (001) B-layers close to pure SrO planes. Furthermore, away from the domain walls, neighboring down (or up) *z*-chains mostly possess dipoles that are more-or-less constant in direction and magnitude within a given *(001)* B-layer (note, however, that the ``topological defect'' regions circled in Fig. 3(b) constitute an exception to that directional trend). Such features combined with the fact that the in-plane dipoles vary in direction when moving along *z* within a given *z*-chain results in complex curvatures in the domain patterns [see, e.g., the curved lines in Figs. 3(b) and (f)]. It is important to note that (i) curvature has been proposed as the driving mechanism of GF in cholesteric blue phases[14,27], (ii) curvature leads to multiple frustrated orientational states, and, (iii) cooperative phenomena and complex pattern formation are characteristic hallmarks of GF[7]. Cooperative dipolar arrangements also lead to Herringbone-type domain patterns[28-30] that are elongated along the *x*-axis [see, Figs. (3b,3f)]. The angle formed by the Herringbone's kinks depend on the position of the (001) B-layer (consistent with Fig. 3e). Each kink of the Herringbone lies on a domain wall. Interestingly, Herringbone patterns have also been reported in frustrated systems[17].

Regarding microscopic characteristics of Phase I, Figure 4 reveals that the individual *z*-chains can still exhibit a variation of their in-plane dipoles along the growth direction [see Fig. 4(b)], but at a much smaller extent than the *z*-chains of Phase II – as a result of the fact that compressive strain desires to annihilate in-plane components of dipoles in favor of enhancing their *z*-components[25]. A striking feature of Phase I is that the domain walls separating ``up'' domains from ``down'' domains can move between adjacent *(010)* planes [see, e.g. Fig. 4(d) *versus* Fig. 4(e)] and that the width of the ``up'' and ``down'' domains can even change between neighboring *(010)* planes [see, e.g. Fig. 4(f) *versus* Fig. 4(g)]. Interestingly, the same features occur between adjacent *(100)* planes, therefore leading to the propagation of two-dimensional waves (connecting the center of gravity of neighboring domain walls) inside the compositionally graded system. Such spatial modulation is reminiscent to that obtained in magnetism[16], wherein frustration effects, when sufficiently strong, stabilize spiral states. The presently discovered dipolar arrangements permit the domain walls to evolve through various representative domain configurations, all of which are equally probable which explains the degeneracy of the exotic states associated with Phase I. Moreover, the novel spatial arrangements of atoms depicted in Figs. 4, automatically lead to cancellation of both the out-of-plane and in-plane components of the polarization in Phase I that is the sum all the dipoles' vectors over the whole graded system vanishes.

A possible microscopic origin for GF to occur in the studied compositionally modulated ferroelectric system may reside in the fact that each *(001)* B-layer possesses its own magnitude for their *z*-dipoles, as a result of the change in Ba composition happening along the *[001]* pseudo-cubic direction [see Figs. 1(e) and 1(f)]. This difference-in-magnitude between adjacent *(001)* B-layers prevents the formation of states that are homogeneously polarized along *[001]*, because such formation would lead to costly-in-energy depolarizing fields (originating from the dipoles being inhomogeneous between different *(001)* B-layers). It also prevents the occurrence of the ``regular'' periodic nanostripe domains found in BT films or long-period BT/ST superlattices[25] because the difference in magnitude between the *z*-dipoles of any two adjacent *(001)* layers is costly in terms of short-range dipolar interactions (and thus adds up to the cost of short-range energy associated with the formation of domain walls). The composition modulation along the growth direction therefore prevents the formation of ``regular'' polarized or domain states, because it dramatically affects long-range interactions that are related to depolarizing fields and short-range interactions. Such composition modulation thus naturally leads to frustrated systems, as long as the *z*-component of the dipoles exists (consistent with the fact that we did not find any sign of frustration for large tensile stains in the corresponding *aa* ground-state). Comparing the different energetic terms of the used effective Hamiltonian approach between ``regular'' nanostripe domains and our reported microstructures also reveals that inhomogeneous strains play a key role in the formation of the microstructures and rich textures associated with Phases I and II, as analogous with the finding of Ref. [30] about the origins of some complex states. All the anomalous features described here (that are consistent with GF) were also numerically found in another BST modulated system. In that latter compound, the composition continuously changes by 25% along the growth direction (rather than by 12.5%, as studied here and as indicated in Fig. 1a).

Our results thus strongly suggest that compositionally graded ferroelectric systems form a novel class of materials that can exhibit geometric frustration. They can therefore provide a novel route to study characteristics of GF, since compositionally graded ferroelectrics can be experimentally realized[18,24] via, e.g., molecular beam epitaxy and pulsed laser deposition. Future studies could also examine the influence of static and dynamic electric fields or defects on the behavior of these modulated systems, which may be relevant for understanding the physics of relaxors and glasses and may have implications in diverse fields.

**Methods**

The total internal energy, $E_{tot}$, of the studied BST graded system is provided by the effective Hamiltonian approach of Ref. [26], with its parameters having been determined from first-principle calculations on relatively small supercells. This effective Hamiltonian scheme has been shown to accurately reproduce the experimental composition-temperature phase diagram of disordered BST solid solutions[26], as well as, to provide predictions that are in good agreement with *ab-initio* calculations in $BaTiO_3/SrTiO_3$ superlattices[25]. The degrees of freedom for $E_{tot}$ are the homogeneous and inhomogeneous strains[26], and the local soft modes $u_i$ in all 5-atom unit cells *i*. Note that $u_i$ is directly proportional to the electric dipole in the cell *i*, and is centered on the Ti-sites unlike the alloying that occurs on the A sites of the $ABO_3$ perovskite structure in BST. The Ba and Sr atoms are randomly distributed in each *(001)* AO plane, with the overall Ba and Sr compositions of each AO plane following the saw-like modulation depicted in Fig. 1a. The epitaxial growth of the investigated modulated BST material on a given substrate is mimicked by freezing some components of the homogeneous stain tensor[22,25,26], in order to impose that each *(001)* layer has the same in-plane lattice constant than the chosen substrate[25]. The zero in misfit strain corresponds to the predicted lattice constant of the cubic disordered $Ba_{0.5}Sr_{0.5}TiO_3$ alloy interpolated down to 0K[26]. The $E_{tot}$ energy is used in Monte-Carlo simulations (with up to 300,000 sweeps) on 12x12x16, 16x16x16 or 20x20x16 supercells to obtain finite-temperature properties, such as the misfit-versus-strain phase diagram and static dielectric responses. The low-temperature microstructures are obtained via gradual cooling (in 5K steps) from 1000K, iteratively reading in configurations from simulated results for the previous temperature.

**Competing interest statement** The authors declare that they have no competing financial interests.

**Correspondence** and requests for materials should be sent to N.C. (narayani@uark.edu).

**Acknowledgements** This work was supported by the National Science Foundation, the Office of Naval Research and the Department of Energy. We gratefully acknowledge extensive use of the supercomputing resources of the University of Arkansas High Performance Computing Center as well as the Center for Piezoelectrics by Design, College of William and Mary, VA. We thank Amy Apon, David Chaffin, Jeff Pummill and Eric J. Walter for computational support.

**Author contributions** This work is an outgrowth of an ongoing project on compositionally modulated ferroelectrics at the University of Arkansas. L.W., S.L. and L.B. developed an effective Hamiltonian implementation for BST systems. N.C. carried out the present Monte-Carlo simulations using these effective Hamiltonian and code implementations. N.C. found exotic degenerate ground states and spiral domains and suggested that these complex results can be explained in terms of geometric frustration. L.B. proposed additional studies of critical behaviors and size dependency and these additional simulations were carried out by N.C. Various complex details were jointly analyzed by N.C. and L.B. and they together wrote the paper, with feedbacks from L.W. and S.L.

**Figure Captions**

**Figure 1**: **Characteristics and material properties. a**, The composition modulation along the growth [001] direction. **b,** The computed phase diagram. Typical error bars are 0.4% misfit strain (lateral) and 150K (vertical). **c-d,** Temperature dependence of the static dielectric susceptibility and its log-log plot (inset) in Phase I and II, respectively. The red solid lines represent the fit of the dielectric susceptibility by the power law $1/\chi \propto (T-T_c)^{\gamma}$ for $T>T_c$. The possible range of values of the $\gamma$ exponent for Phases I and II are indicated in panel (b). **e-f,** Layer-by-layer average of the dipoles' magnitude and Cartesian components.

**Figure 2**: **Ground state microstructures**. **a-f,** Top view of various ground state microstructures associated with Phase II (a-c) and Phase I (d-f) at T=20 K. Red and blue colors refer to the *z*-component of the dipoles $u_z$ being positive and negative, respectively. The exotic microstructures of Panels (a-c) (respectively, (d-f)) have similar internal energy. The critical exponent $\gamma$ is indicated in each panel. Selected flipped sites occurring near the domain walls are highlighted in Panel (a). Different supercell sizes are used for the same misfit strain (for instance, panels (d), (e) and (f) correspond to 12x12x16, 16x16x16, 20x20x16 supercells, respectively).

**Figure 3**: **Frustrated ground state microstructure and dipolar orientational correlations in Phase II. a,** A ground state dipolar configuration. Red and blue colors refer to $u_z>0$, $u_z<0$, respectively. **b,** Bending of lines, curvature and Herringbone patterns in the microstructure are indicated via solid black lines. Selected z-chains are highlighted and some ``topological defects'' are circled. **c-e**, Orientational correlation functions (see text). The blue and red curves in panels (c)-(e) indicate that the first dipole (used to compute these correlation functions) belongs to a up or down *z*-chain, respectively. **f,** The dipoles forming the Herringbone (black lines in panel b).

**Figure 4**: **Frustrated ground state microstructure and spiral states in Phase I. a,** An example of pattern of dipoles for a possible frustrated ground state belonging to Phase I (misfit strain of -2.6%) at 20K. Panel (a) displays a three-dimensional view, which reveals spiral arrangements. **b,** A single *z*-chain (see text). **c,** an assembly of *z*-chains being adjacent along the [010] pseudo-cubic direction. **d-g,** The dipolar arrangement in different, successive (010) planes. Red and blue colors refer to *z*-component of the dipoles being positive and negative, respectively.

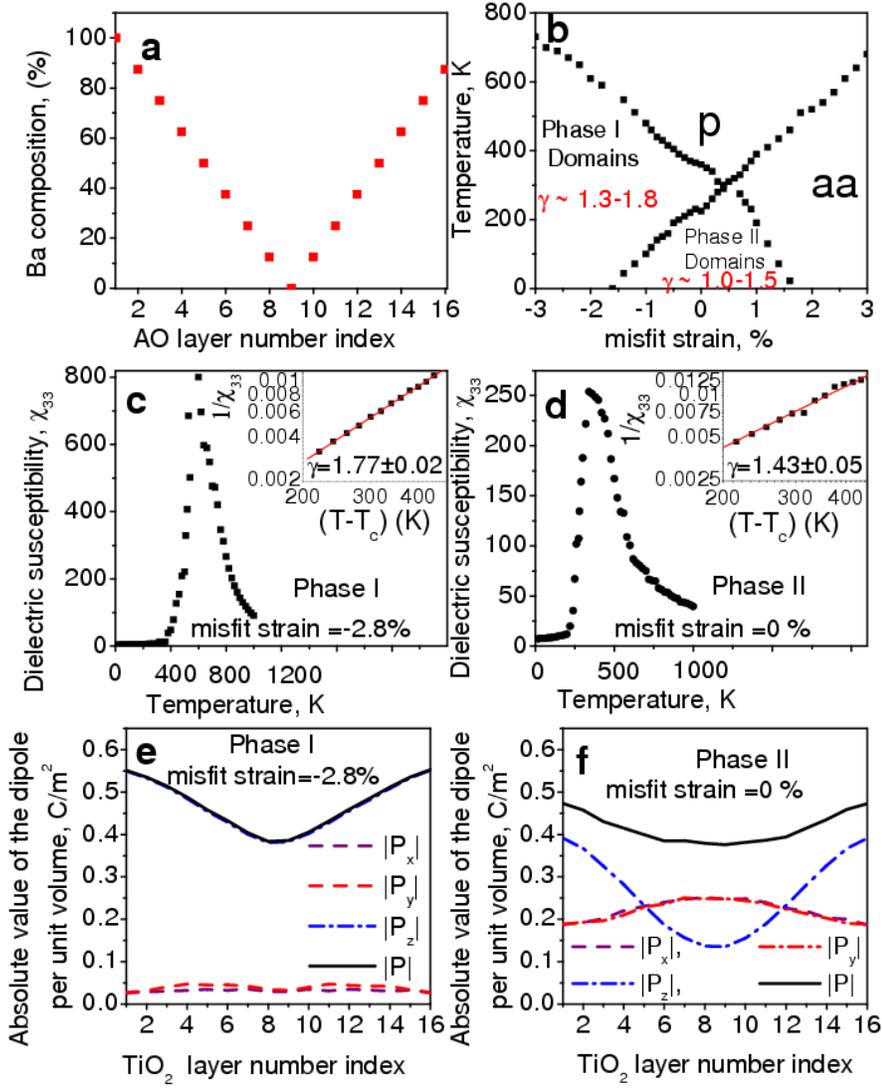

Figure 1

**Figure 1**: **Characteristics and material properties. a,** The composition modulation along the growth [001] direction. **b,** The computed phase diagram. Typical error bars are 0.4% misfit strain (lateral) and 150K (vertical). **c-d,** Temperature dependence of the static dielectric susceptibility and its log-log plot (inset) in Phase I and II, respectively. The red solid lines represent the fit of the dielectric susceptibility by the power law $1/\chi \propto (T-T_c)^\gamma$ for $T>T_c$. The possible range of values of the $\gamma$ exponent for Phases I and II are indicated in panel (b). **e-f,** Layer-by-layer average of the dipoles' magnitude and Cartesian components.

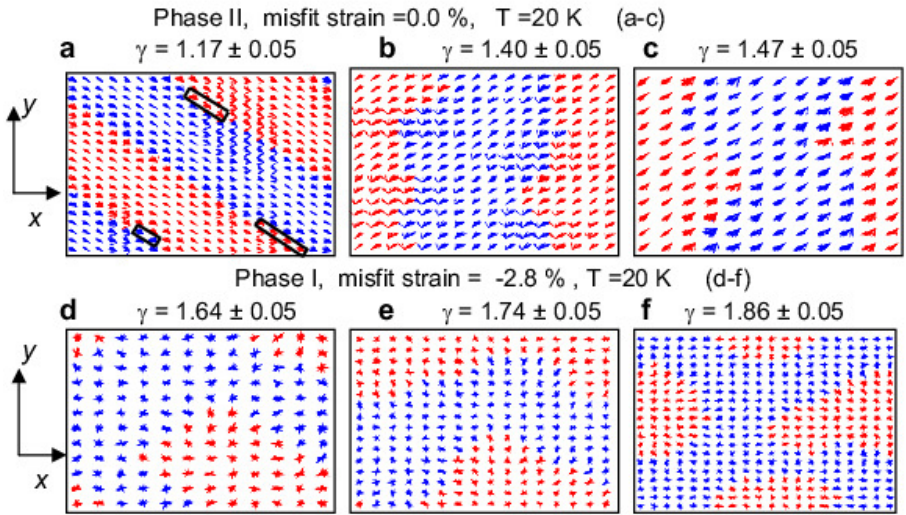

Figure 2

**Figure 2**: **Ground state microstructures**. **a-f,** Top view of various ground state microstructures associated with Phase II (a-c) and Phase I (d-f) at T=20 K. Blue and red colors refer to the *z*-component of the dipoles $u_z$ being positive and negative, respectively. The exotic microstructures of Panels (a-c) (respectively, (d-f)) have similar internal energy. The critical exponent γ is indicated in each panel. Selected flipped sites occurring near the domain walls are highlighted in Panel (a). Different supercell sizes are used for the same misfit strain (for instance, panels (d), (e) and (f) correspond to 12x12x16, 16x16x16, 20x20x16 supercells, respectively).

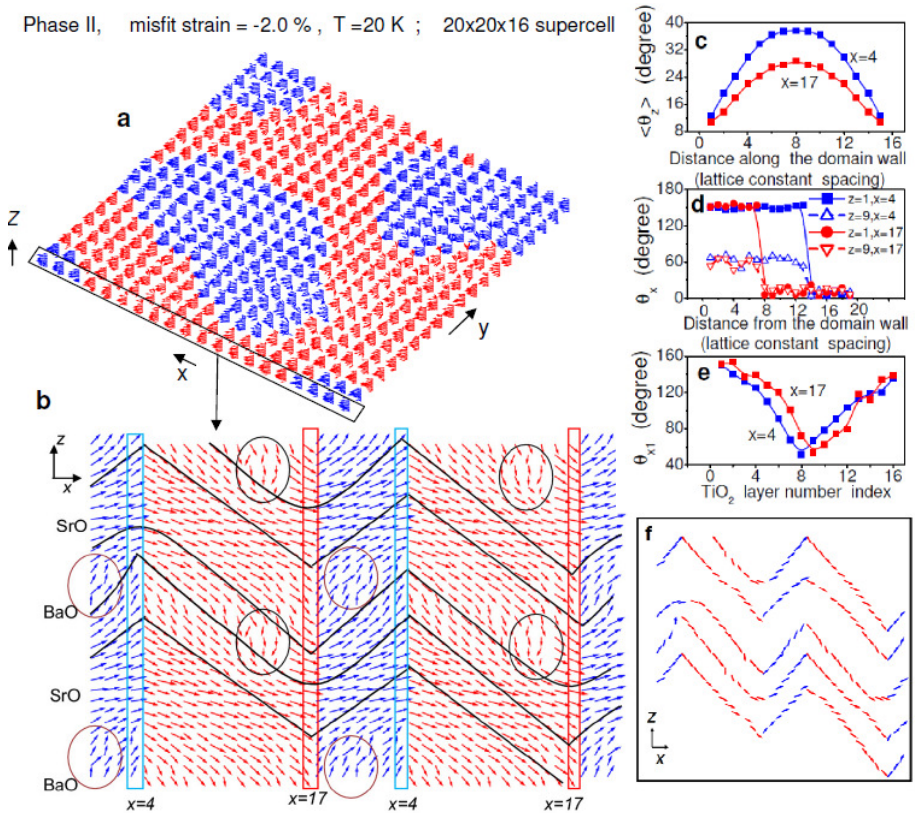

Figure 3

**Figure 3**: **Frustrated ground state microstructure and dipolar orientational correlations in Phase II. a,** A ground state dipolar configuration. Blue and red colors refer to $u_z>0$, $u_z<0$, respectively. **b,** Bending of lines, curvature and Herringbone patterns in the microstructure are indicated via solid black lines. Selected z-chains are highlighted and some ``topological defects'' are circled.   **c-e**, Orientational correlation functions (see text). The blue and red curves in panels (c)-(e) indicate that the first dipole (used to compute these correlation functions) belongs to a up or down *z*-chain, respectively. **f,** The dipoles forming the Herringbone (black lines in panel b).

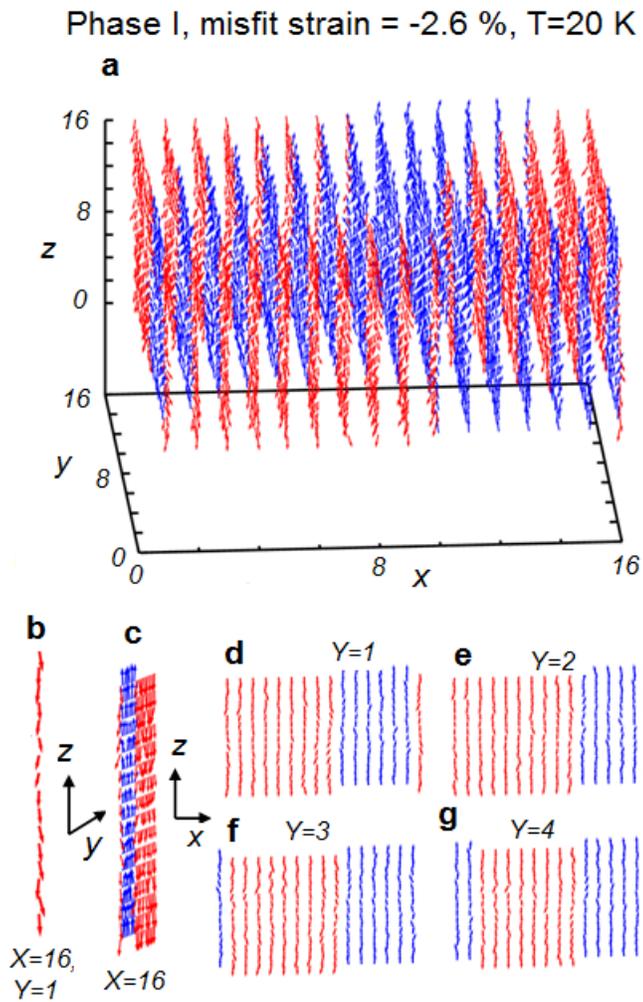

Figure 4

**Figure 4**: **Frustrated ground state microstructure and spiral states in Phase I. a,** An example of pattern of dipoles for a possible frustrated ground state belonging to Phase I (misfit strain of -2.6%) at 20K. Panel (a) displays a three-dimensional view, which reveals spiral arrangements. **b,** A single $z$-chain (see text). **c,** an assembly of $z$-chains being adjacent along the [010] pseudo-cubic direction. **d-g,** The dipolar arrangement in different, successive (010) planes. Blue and red colors refer to $z$-component of the dipoles being positive and negative, respectively.

# Geometric frustration in compositionally modulated ferroelectrics

Narayani Choudhury[1], Laura Walizer[2], Sergey Lisenkov[3] & L. Bellaiche[1]

[1]Department of Physics, University of Arkansas, Fayetteville, Arkansas 72701, USA.

[2]Engineer Research and Development Center, Vicksburg, Mississippi 39180, USA.

[3]Department of Physics, University of South Florida, Tampa, Florida 33620, USA.

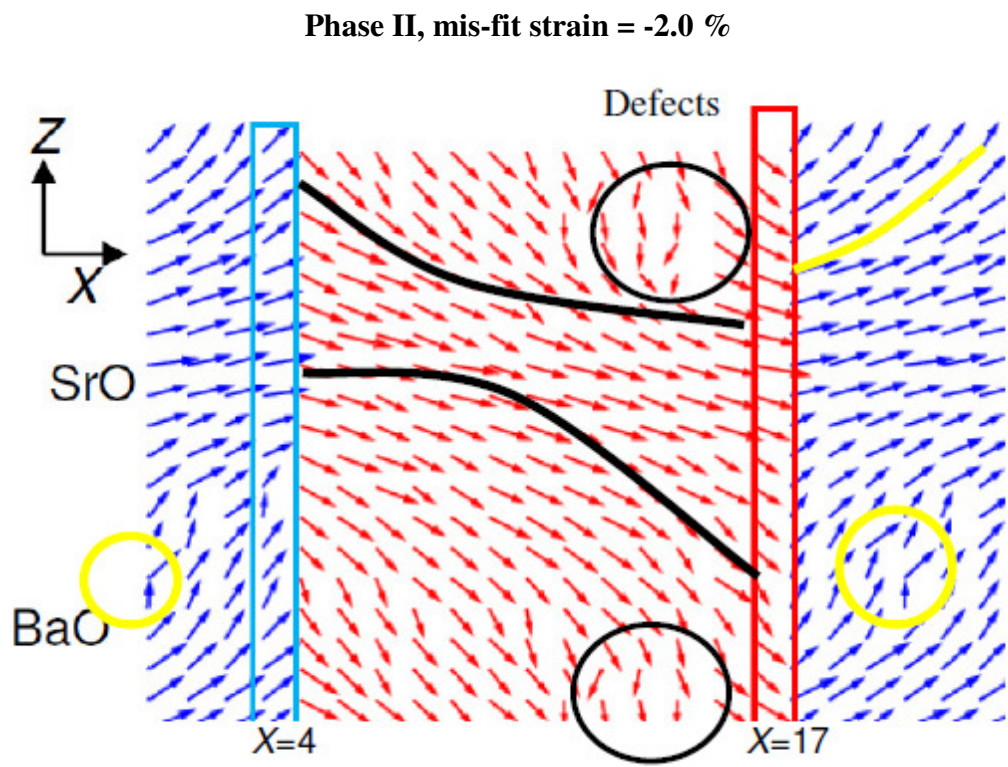

**Figure S1**: **Bending of dipolar lines and curvatures:** Electric dipoles bending (shown as black and yellow lines) and curvatures in compositionally graded ferroelectrics. The dipole patterns shown are in a given (010) plane. Selected topological defects are circled and some domain walls are highlighted. Blue, positive dipole $z$ component; red, negative dipole $z$ component. Additional details are given in figure 3 of the paper.